\documentclass[11pt]{article}

\usepackage{amsmath}
\usepackage{amssymb}
\usepackage{graphicx}
\usepackage{cite}
\usepackage[T1]{fontenc}
\usepackage{textcomp}
\usepackage{mathcomp}

\newenvironment{minilinespace}{
	\baselineskip = 1mm
}

\usepackage[labelfont=bf,labelsep=period,justification=raggedright]{caption}

\makeatletter
\renewcommand{\@biblabel}[1]{\quad#1.}
\makeatother

\begin{document}

\begin{flushleft}
{\Large {\sf
\textbf{WARN: an R package for quantitative reconstruction of weaning ages in archaeological populations using bone collagen nitrogen isotope ratios}
}}
\vspace{5mm}

\sf{Takumi Tsutaya$^{1,\ast}$, Minoru Yoneda$^{1,2}$}
\vspace{5mm}

{\small {\sf $^{1}$ Department of Integrated Biosciences, Graduate School of Frontier Sciences, The University of Tokyo, Kashiwanoha 5-1-5, Kashiwa, Chiba, Japan. $^{2}$ University Museum, The University of Tokyo, Hongo 7-3-1, Bunkyo, Tokyo, Japan.}}
\vspace{5mm}

{\small {\sf $^{\ast}$ Corresponding author; E-mail: 127309@ib.k.u-tokyo.ac.jp}}
\end{flushleft}

\noindent \rule{\textwidth}{0.1mm}
\section*{{\sf Abstract}} 
\par {\sf Nitrogen isotope analysis of bone collagen has been used to reconstruct the breastfeeding practices of archaeological human populations. However, weaning ages have been estimated subjectively because of a lack of both information on subadult bone collagen turnover rates and appropriate analytical models. Here, we present a model for analyzing cross-sectional $\rm{{\sf \delta^{15}N}}$ data of subadult bone collagen, which incorporates newly estimated bone collagen turnover rates and a framework of approximate Bayesian computation. Temporal changes in human subadult bone collagen turnover rates were estimated anew from data on tissue-level bone metabolism reported in previous studies. A model for reconstructing precise weaning ages was then developed and incorporating the estimated turnover rates. The model is presented as a new open source R package, WARN (Weaning Age Reconstruction with Nitrogen isotope analysis), which computes the age at the start and end of weaning, $\rm{{\sf ^{15}N}}$-enrichment through maternal to infant tissue, and $\rm{\sf{\delta^{15}N}}$ value of collagen synthesized entirely from weaning foods with their posterior probabilities. A precise reconstruction of past breastfeeding and weaning practices over a wide range of time periods and geographic regions could make it possible to understand this unique feature of human life history and cultural diversity in infant feeding practices.}

\noindent \rule{\textwidth}{0.1mm}

\section*{Introduction}
\par Investigating variations in the breastfeeding and weaning practices of ancient human populations can provide information on the health, cultural traits, and reproduction of these populations. Breast milk provides various antibodies as well as nutrition to infants, and is important for subadult survival \cite{Cunningham1995, Kramer2004}. Breastfeeding practices are closely related to the growth of subadults and overall health of a population \cite{Dettwyler1992, Katzenberg1996, Lewis2007}. The type of subsistence activities, social constructs, diet and various cultural factors affect breastfeeding practices \cite{Fildes1995, Ford1964, Maher1992}, and the length of the breastfeeding period is one of the most important determinants of the fertility of a population \cite{Bongaarts1978, Bongaarts1982}. Shorter breastfeeding periods tend to result in shorter birth intervals, and, in turn, higher fertility because breastfeeding can delay the resumption of ovulation \cite{Wood1994, Ellison1995, McNeilly2001, Valeggia2009}. Furthermore, it is supposed that humans are weaned earlier than the other great apes, and understanding evolutionary changes in weaning practices is of great interest \cite{Bogin1997, Hawkes1998, Humphrey2010, Kennedy2005, Lee1996, Sellen2007}.

\par Stable isotope analysis of bone collagen is useful for reconstructing the dietary habits of ancient people, and it has also been used to reconstruct breastfeeding and weaning practices of archaeological populations \cite{Dittmann2000, Fogel1989, Herring1998, Richards2002, White1994}. Nitrogen isotope ratios ($\rm{\delta^{15}N}$ values) of body proteins primarily reflect dietary protein isotope ratios \cite{Ambrose1993, Tieszen1993}. Prior to and immediately after birth, $\rm{\delta^{15}N}$ values of infants are the same as those of their mothers \cite{Fuller2004}. After birth, infants who are exclusively breastfed show 2--3\textperthousand\ higher $\rm{\delta^{15}N}$ values than their mothers \cite{Fogel1989, Fuller2006} because of the trophic level effect \cite{Bocherens2003, Minagawa1984, Schoeninger1984}. Subadult $\rm{\delta^{15}N}$ values decrease after the introduction of supplementary foods, and gradually approach the values found in adult bone collagen. It is possible to reconstruct infant feeding practices of an archaeological populations by combining $\rm{\delta^{15}N}$ values and physically estimated ages at death of subadults of different ages \cite{Lewis2007, Scheuer2000}.

\par However, in previous isotopic studies, weaning ages have been subjectively estimated from visual assessments of detectable changes in subadult bone collagen $\rm{\delta^{15}N}$ values. To overcome these difficulties, attempts have been made to simulate changes in $\rm{\delta^{15}N}$ values of subadult bone collagen in two pioneering studies. Schurr \cite{Schurr1997} used exponential functions to describe changes in $\rm{\delta^{15}N}$ values and estimate the age at the start of weaning. Millard \cite{Millard2000} suggested that the model proposed by Schurr \cite{Schurr1997} suffered from a number of difficulties, and proposed an alternative model that further included a nitrogen mass balance and the age at the end of weaning. However, both models still suffer from the following three problems.
\begin{enumerate}
	\item{The subadult bone collagen turnover rates are not fully considered. The bone collagen turnover rate is high in early infancy \cite{Bryant1964, Rivera1965}, but it decreases over the course of subadult growth \cite{Hedges2007, Szulc2000}. If not corrected, the lower bone collagen turnover rates at higher ages would generate significant discrepancies between the visible changes in bone $\rm{\delta^{15}N}$ values and actual weaning ages.}
	\item{Some parameters used to describe changes in $\rm{\delta^{15}N}$ values are determined arbitrarily. Two parameters, $\rm{^{15}N}$-enrichment from maternal to infant tissues and the $\rm{\delta^{15}N}$ values in weaning foods, could vary among different individuals and populations; therfore, they should be considered as variables in addition to the weaning ages. First, it has been reported that $\rm{^{15}N}$-enrichment varies to some extent in modern infant-mother pairs (between 1.7\textperthousand\ and 2.8\textperthousand, $n = 7$: \cite{Fuller2006}) and in archaeological populations (between 0.5\textperthousand\ and 4.4\textperthousand, $n = 25$: \cite{Waters-Rist2010}). Second, it is possible that $\rm{\delta^{15}N}$ values of materials used in weaning foods were different than those used in adult foods \cite{Dupras2001, Keenleyside2009}.}
	\item{The results are represented as point estimates without either probabilities or confidence intervals. The probabilities of the weaning parameters should be calculated to evaluate the validity of the computation results.}
\end{enumerate}

\par The objective of this study is to develop a model for analyzing cross-sectional $\rm{\delta^{15}N}$ data of subadult bone collagen in archaeological skeletal populations. The model is programmed in R language, which is a free software environment for statistical computing and graphics \cite{R2012}. The model has the following three important features that are not present in the previous models:
\begin{enumerate}
	\item{The subadult bone collagen turnover rate is estimated anew  and incorporated in the equations.}
	\item{The enrichment factor and $\rm{\delta^{15}N}$ values of weaning foods are included as target parameters to be estimated.}
	\item{Using a framework of approximate Bayesian computation (ABC) allows researchers to calculate the probabilities and credible intervals of the weaning parameters.}
\end{enumerate}

\subsection*{Subadult bone collagen turnover rate}
\par
\begin{figure}[!ht]
\begin{center}
	\includegraphics{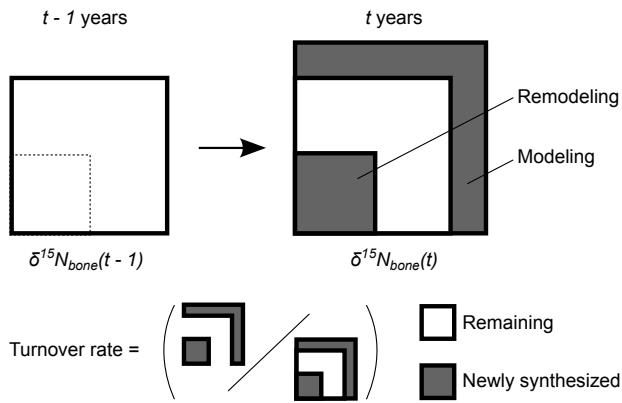}
\end{center}
\caption{{\bf Schematic illustration of the bone turnover process.} The $\rm{\delta^{15}N}$ value for bone collagen at the unit time age of $t$ years is represented as $\delta^{15}N_{bone}(t)$.}
\label{schematic}
\end{figure}

 Temporal changes in the bone collagen turnover rate must be considered to estimate a precise weaning ages from an observed isotope ratio. Bone collagen is laid down during childhood because of bone modeling, which is a formative process primarily associated with skeletal growth, and is replaced throughout life by bone remodeling, which is a coupled resorptive and formative process that does not change the quantity of bone \cite{Fratzl2004, Glimcher2006}. As indicated in Figure \ref{schematic}, turnover refers to the proportion of newly synthesized bone collagen to the total bone collagen during modeling and remodeling over a unit of time. When the turnover rate is high enough (i.e., $\geq$1.0 per unit time), bone collagen at a specific age consist only of newly synthesized collagen, and the isotope ratio will immediately change with dietary changes. When the turnover rate is lower (i.e., $<$1.0), the bone collagen consists not only of newly synthesized but also previously synthesized collagen, the isotope ratio reflects recent and past dietary intakes.

\par Although temporal changes in turnover rates of subadult bone minerals and collagen have been estimated by analyzing the uptake of $\rm{^{90}Sr}$ fallout \cite{Rivera1965, Papworth1984} and bomb-$\rm{^{14}C}$ \cite{Hedges2007}, respectively, the estimates produced from these bulk cross-sectional studies were not necessarily precise. Some of the estimates were not based on direct measurements in subadults but on extrapolations from results for adults. In addition, the assumptions made about the dietary intake of tracers in these subadults were simplistic and ignore individual variation. In the present study, we calculated turnover rates from bone metabolism mechanisms at the tissue level so that more precise subadult bone turnover rates could be estimated.

\par Turnover rates of mineral and organic phases should differ because the mineralization process is much slower than the synthesis of the organic matrix. Bone is a composite material, and is mainly made of a calcified organic matrix \cite{Glimcher2006, Rho1998}. The microstructure of bone material consists of assembled collagen fibrils forming the organic phase, and tiny mineral particles reinforcing them \cite{Fratzl2004}. Two coupled processes are responsible for bone remodeling. The resorptive process involves osteoclasts dissolving the mineral phase by creating a low pH environment around the bone surface, and then producing a lysosomal protease to degrade the organic matrix \cite{Teitelbaum2000}. The next formative process inolves osteoblasts replacing the organic matrix and rapidly mineralizing it to up to 70\% of full mineralization capacity within a few days (primary mineralization), the residual 30\% of the mineralization occurring gradually over several years (secondary mineralization) \cite{Fratzl2004, Ruffoni2007}. The mineralization process has been formulated as a mineralization law \cite{Ruffoni2007}. Bone modeling occurs with a similar formative process as bone remodeling but with the resorption of the bone cartilage template instead of mineralized old bone \cite{Scheuer2000}. Since the growth \cite{Mitchell1945} and replacement \cite{Leggett1982} (i.e., turnover, consisting of the modeling and remodeling processes) of bone minerals at the tissue level have been well documented, the turnover of bone collagen can be estimated by correcting the mineralization delay \cite{Ruffoni2007}.

\subsection*{Approximate Bayesian computation}
\par ABC is a modern approach in Bayesian inference that allows posterior distributions to be evaluated when it is difficult to calculate the likelihood function, which describes probabilities under given parameters. Various ABC methods have been applied in diverse fields such as population genetics, evolutionary biology, ecology, and epidemiology \cite{Beaumont2010, Bertorelle2010, Csillery2010}.

\par A general ABC algorithm takes a given observation $x$ and repeat the following three steps until $J$ points have been accepted:
\begin{enumerate}
	\item Draw the candidate parameter $\theta_j$ from the prior distribution $\pi(\theta)$.
	\item Simulate dataset $x_j$ using $\theta_j$ and the model.
	\item Accept $\theta_j$ if $\rho(x, x_j) \leq \alpha$, and otherwise reject $\theta_j$.
\end{enumerate}
Here $\rho(\cdot)$ is a function measuring the distance between simulated and observed data points, $\alpha$ is a fixed tolerance for the ``closeness'' of simulated and observed data, and $x$, $x_j$, and $\theta$ may be vector values. If $\rho(\cdot)$ measures appropriate distances and tolerance is sufficiently small, the accepted parameters reasonably approximate the posterior distributions. This is a rejection sampling algorithm, which is the simplest ABC procedure.

\par Although ABC has proved to be a flexible and powerful approach for evaluating posterior distributions, its major drawback is its inefficiency. Acceptance rates in the simple rejection sampling described above can be very low, especially when the posterior is a long way from the prior, which wastes computing time. Several algorithms have been proposed to increase the sampling efficiency, by introducing weighting with regression analysis \cite{Beaumont2002, Leuenberger2010}, Markov chain Monte Carlo sampling \cite{Marjoram2003}, and sequential Monte Carlo (SMC) sampling \cite{Beaumont2009, Sisson2007, Toni2009}. We used SMC sampling with corrected partial rejection control proposed by Sisson et al. \cite{Sisson2007} because this method could be implemented more quickly and simply in our model in the R software environment. SMC sampling is characterized by successively decreasing the tolerance, and weighted resampling from the previous parameter population.

\section*{Materials and Methods}
\subsection*{Estimating subadult bone collagen turnover rates}
\par In this study, bone collagen turnover rates in subadults were calculated from the modeling \cite{Mitchell1945} and remodeling \cite{Leggett1982} rates for cancellous bone minerals, and the mineralization law for the bone organic matrix \cite{Ruffoni2007}. ``Turnover'' is defined as the aggregated effects of bone modeling (i.e., the addition of bone tissue by skeletal growth) and remodeling (i.e., the replacement of existing bone tissues). First, following Leggett et al. \cite{Leggett1982}, the bone mineral turnover rate $T_{min}[t]$ over one unit of time (i.e., one year from $t-1$ to $t$, Equation 4) in childhood was calculated using the functions that describe the temporal change in bone mineral mass $C(t)$ \cite{Mitchell1945} (Equation 1) and the remodeling rate $\gamma(t)$ \cite{Leggett1982} (Equation 2). Next, the bone collagen turnover rates $T_{col}[t]$ over one unit of time from $t - 1$ to $t$ years (Equation 5) were calculated sequentially with $T_{min}[t]$ and the mineralization law, $\lambda(i)$, which described the bone collagen mineralization process \cite{Ruffoni2007}. The mineralization law was derived from Ruffoni et al. \cite{Ruffoni2007}, and represents the rate of mineralization of the collagen portion at the $i$th year after the collagen matrix was formed (Equation 3). Finally, the resulting discrete turnover rates were coerced into a quartic polynomial (QP) formula (Equation 6). Turnover rates at ages less than one year were extrapolated from the QP function.

\par Basic functions to describe the temporal changes of bone mineral were derived from several previous studies. Following Mitchell et al. \cite{Mitchell1945}, the bone mineral mass $C(t)$ at age of $t$ years was represented as:\\
	$C(t) = 28.0 + 86.828t - 16.5105t^2 + 1.5625t^3 - 0.04114t^4$	($0 \geq t \geq 20$) (Equation 1).\\
This equation represents the modeling process of bone turnover. On the other hand, the remodeling rate $\gamma(t)$ at age of $t$ years was represented as follows:
\begin{enumerate}
	\item{$\gamma(t) = \frac{104.3}{C(t)}$, when $t \leq 1.5$}, and
	\item{$\gamma(t) = 0.975\mathrm{e}^{-0.11t}$, when $t > 1.5$} (Equation 2).
\end{enumerate}
This equation was obtained from Leggett et al. \cite{Leggett1982} and was derived from direct histological observations of subadult rib bone formation and resorption performed by Frost \cite{Frost1969}. Note that a term for the radioactive decay of $\rm{^{90}Sr}$ (0.025 per year), included in the original equations, was excluded from our equations. Following Ruffoni et al. \cite{Ruffoni2007}, the mineralization law $\lambda(i)$, which describes the rate of mineralized collagen portion at $i$th years after the collagen matrix was formed, was set as:\\
	$\lambda(i) = c_1 \frac{1 + \frac{i}{i_1}}{\frac{i}{i_1}} + c_2 \frac{1 + \frac{i}{i_2}}{\frac{i}{i_2}}$ (Equation 3).\\
In this study, $c_1$, $i_1$, $c_2$ and $i_2$ were set as $18/23$, $1/300$, $25/92$, and $5$, respectively. Equation 3 corresponds to over 70\% primary mineralization in a few days and protracted secondary mineralization of up to 100\% in about 20 years, values that have been given in several previous studies \cite{Akkus2003, Fratzl2004, Ruffoni2007, Ruffoni2008}.

\par Temporal changes in the bone mineral turnover rate can be represented using these functions. Put simply, the bone mineral turnover rate, $T_{min}[t]$, over one unit of time from $t - 1$ to $t$ years, was represented as follows:\\
	$T_{min}[t] = \frac{C(t) - C(t - 1)}{C(t)} + \int_{t-1}^{t}\gamma(x)dx \frac{C(t - 1)}{C(t)}$ (Equation 4).\\
The former and latter terms in the function indicate the effects of bone modeling and remodeling, respectively. Using the bone collagen turnover rate, $T_{col}[t]$, over one unit of time from $t - 1$ to $t$ years, $T_{min}[t]$ can also be represented as follows:
\begin{enumerate}
	\item{$T_{min}[t] = T_{col}[t] \Delta\lambda[1]$, when $t = 1$, and}
	\item{$T_{min}[t] = T_{col}[t] \Delta\lambda[1] + \sum_{j = 1}^{t - 1}(T_{col}[j] \frac{C(j)}{C(t)} \Delta\lambda[t + 1 - j])$, when $t \geq 2$ (Equation 5).}
\end{enumerate}
The former and latter terms in the second function shown in Equation 5 indicate the effects of turnover delay in the bone mineral for the intended unit of time (i.e., $t-1$ to $t$ years) and the aggregated effects of the delay for the former unit times (i.e., $0$ to $1$ year, $1$ to $2$ years, .., and $t-2$ to $t-1$ years). The turnover rate over one unit of time (i.e., one year) from $t - 1$ to $t$ years can be sequentially calculated using Equation 4 and 5. The resulting discrete turnover rates were coerced to a quartic polynomial formula using the \textit{nls} function in R, in accordance with the quartic formula for bone mineral mass (i.e., Equation 1). The formula is represented as follows:\\
	$T_{col}[t] = 1.778 - 0.4121t + 0.05029t^2 - 0.002756t^3 + 0.0005325t^4$ (Equation 6).

\subsection*{Changes in $\rm{\delta^{15}N}$ values of diet and bone collagen}
\par Following Millard \cite{Millard2000}, the $\rm{\delta^{15}N}$ value of newly synthesized collagen at a given age of $t$ years was defined by four parameters, the ages at the start ($t_{1}$) and end ($t_{2}$) of weaning, enrichment factor between the infant and mother ($E$), and $\rm{\delta^{15}N}$ value of collagen synthesized entirely from weaning foods ($\delta^{15}N_{wnfood}$) (Equation 7 and 8). The $\rm{\delta^{15}N}$ value of newly synthesized collagen equals the sum of the $\rm{\delta^{15}N}$ value of the mothers tissue and enrichment factor before weaning ($t < t_{1}$), which changes exponentially during weaning ($t_{1} \leq t \leq t_{2}$), and equals the collagen $\rm{\delta^{15}N}$ value that fully reflects the consumption of supplementary food ($t > t_{2}$). Then, the incorporation of newly synthesized collagen and replacement of existing collagen in bone are simulated in over each successive unit time using the estimated turnover rate for bones (Equation 6 and 9). As most isotopic studies on weaning have focused on rib bones, because of their assumed fast turnover \cite{Parfitt2002} and relatively trivial importance in morphological studies, the rate incorporated into the present model was that of cancellous bones. Although the rib bones that were sampled would have contained cortical parts, the relatively high surface to volume ratio in ribs would have resulted in a high proportion of cancellous parts and only thin cortical parts, making the turnover rate comparable to that of cancellous bones \cite{Parfitt2002}. Although one unit of time consists of one year, adjustments from the last unit of time enables simulated $\rm{\delta^{15}N}$ values to be calculated for each individual in the dataset (Equation 10 and 11). Simulated $\rm{\delta^{15}N}$ values, $\delta^{15}N_{bone}$, for each individual can be calculated under the given weaning parameters ($t_{1}$, $t_{2}$, $E$ and $\delta^{15}N_{wnfood}$) using the model described above. The most appropriate weaning parameters can be estimated by minimizing the mean least square distance between the observed and resultant simulated change in bone collagen $\rm{\delta^{15}N}$ values.

\par Following Millard \cite{Millard2000}, the $\rm{\delta^{15}N}$ values for newly synthesized collagen $\delta^{15}N_{new}(t)$ at a given age of $t$ years are given by the following equation:\\
	$\delta^{15}N_{new}(t) = (1 - p(t)) (\delta^{15}N_{mother} + E) + p(t) \delta^{15}N_{wnfood}$	 (Equation 7).\\
The proportion of non-milk protein in the total dietary protein intake at age of $t$ is represented as $p(t)$. The $\rm{\delta^{15}N}$ value for the mother's milk is described as $\delta^{15}N_{mother} + E$, using the $\rm{\delta^{15}N}$ value for the mothers tissue, $\delta^{15}N_{mother}$ (approximated by the mean $\rm{\delta^{15}N}$ value for adult females), and a $\rm{^{15}N}$ enrichment factor for the transfer from the maternal to infant tissue, $E$. The $\rm{\delta^{15}N}$ value for collagen synthesized from non-milk foods is represented as $\delta^{15}N_{wnfood}$. We considered $\delta^{15}N_{wnfood}$ to be a variable because children in the past could have eaten supplementary foods with different $\rm{\delta^{15}N}$ values from the adult mean $\rm{\delta^{15}N}$ values. This value has been approximated in previous studies as the mean $\rm{\delta^{15}N}$ value for the adults.

\par The proportion of non-milk protein in the total dietary protein intake is assumed, in our model, to increase exponentially. The relative proportion of non-milk protein at the age of $t$ years, $p(t)$, is described as follows:
\begin{enumerate}
	\item{$p(t) = 0$, when $t < t_{1}$ (breast milk only),}
	\item{$p(t) = (\frac{t - t_{1}}{t_{2} - t_{1}})^2$, when $t_{1} \leq t \leq t_{2}$ (during the weaning process), and}
	\item{$p(t) = 1$, when $t > t_{2}$ (no breast milk), (Equation 8),}
\end{enumerate}
where the ages at the start and end of weaning are represented as $t_{1}$ and $t_{2}$, respectively. Equation 8 was derived from a model proposed by Millard \cite{Millard2000} and represents slow initial weaning and rapid final weaning. In the original model, four forms (linear, parabolic, reverse parabolic, and sigmoid) of dietary change were applied to condition 2 in Equation 8. Although the form of dietary change during weaning can be selected in the WARN package, we used only the parabolic form because Millard \cite{Millard2000} used a parabolic weaning pattern to model the changes in $\rm{\delta^{15}N}$ in archaeological datasets. This seems to be a reasonable assumption as not only the amount of milk protein consumed decreases during the weaning process but also the proportion of milk protein consumed also decreases because of the increasing total dietary intake in growing subadults.

\par The $\rm{\delta^{15}N}$ value for bone collagen at the age of $t$ years, $\delta^{15}N_{bone}(t)$, is calculated as follows:\\
	$\delta^{15}N_{bone}(t) = \delta^{15}N_{bone}(t - 1) (1 - T_{col}[t]) + \int_{t - 1}^{1}\delta^{15}N_{new}(t)(x) dx T_{col}[t]$	 (Equation 9).\\
The former and latter parts of the equation represent the remaining and the newly synthesized portion, respectively, of the bone collagen over one unit of time from $t - 1$ to $t$ years. Extending equation 9, the $\rm{\delta^{15}N}$ value for bone collagen at the age of $t + a$, i.e., $a$ being a part of one year from the unit time point $t$ ($0 < a < 1$), is represented as:\\
	$\delta^{15}N_{bone}(t + a) = \delta^{15}N_{bone}(t) (1 - T_{col}[t + a]) + \int_{t}^{t + a}\delta^{15}N_{new}(t)(x) dx T_{col}[t + a]$	 (Equation 10).\\
In equation 10, $T_{col}[T]$ is the bone collagen turnover rate over $a$ year from $t$ to $t + a$, given by:\\
	$T_{col}[t + a] = T_{col}[t + 1] \frac{\int_{t}^{t + a}T_{col}[x] dx}{\int_{t}^{t + 1}T_{col}[x] dx}$	(Equation 11).\\
The bone collagen $\rm{\delta^{15}N}$ values for each unit of time (one year) can be calculated sequentially, as reference values, using Equation 9 under the given parameters. The $\rm{\delta^{15}N}$ values that correspond to the observed ages for the samples can then be calculated from the reference values and Equation 10. The initial bone collagen values at 0 year of age, $\delta^{15}N_{bone}(0)$, were approximated using the mean $\rm{\delta^{15}N}$ value of adult females, because the $\rm{\delta^{15}N}$ value for infant tissue is assumed to be the same as to that of the mother \cite{Fuller2006}. Theoretical $\rm{\delta^{15}N}$ values for the age of each individual in the observed dataset can be calculated using Equation 10.

\par In our model, the differences between the individuals are evaluated by calculating mean square distance, $D$, between the observed and simulated $\rm{\delta^{15}N}$ values. Put simply, point estimates of the parameters with minimized $D$ can be calculated by solving the optimization problem (the application of the optimization problem to palaeo dietary reconstructions has been described by Little and Little \cite{Little1997}). These represent point estimates under the framework of maximum likelihood estimates (MLE). Although the point estimates do not provide information on the error ranges, they will be used later in the SMC sampling procedures; therefore, optimized values for weaning parameters under the MLE framework were calculated. We used the \textit{optim} function in R to obtain the optimized parameter value, $\theta_{opt}$, and its resultant minimum mean square distance, $D_{opt}$.

\subsection*{Incorporation of ABC}
\par To obtain posterior probabilities of the estimated parameters, fitting calculations between the observed and simulated data are performed under the ABC framework with SMC sampling proposed by Sisson et al. \cite{Sisson2007}. Using the ABC framework, a number of weaning parameter sets that give well-fitted $\rm{\delta^{15}N}$ values were sampled and assumed to represent the posterior distributions of the parameters. After applying the ABC procedure, posterior distributions were smoothed using the kernel density estimation \cite{Wand1995}, and joint probabilities for weaning ages ($t_{1}$ and $t_{2}$) and marginal probabilities for $E$ and $\delta^{15}N_{wnfood}$ were calculated. In the density estimation, posterior probabilities were calculated to one decimal places for discrete parameter categories because strictly implementing the density estimation as a continuous distribution requires advanced numerical analysis techniques.

\par SMC sampling is characterized by a successive reduction in tolerance and a weighted resampling from the previous parameter population, called a ``particle''. Particles of preliminary simulations are used to calculate the next set of parameter vectors, to generate simulated data within a certain distance $D$ from the observed data. The particles are then repeatedly resampled (according to a weighting scheme that considers the prior distributions), perturbed (using a transition kernel), and judged (on the basis of a successively decreasing tolerance). The particles after this iterative process finally approximate a sample of the posterior distribution of the parameters. In particular, the partial rejection control procedures prune away parameters that have minimal impacts on the final estimation in the parameter weighting step in the earlier stages of the tolerance reduction, and this increases the sampling efficiency \cite{Liu2001}.

\par To adopt the ABC framework, we added individual error terms $\epsilon_i$ in Equation 10 as follows:\\
		$\delta^{15}N_{bone}(t + a) = \delta^{15}N_{bone}(t) (1 - T_{col}[t + a]) + \int_{t}^{t + a}\delta^{15}N_{new}(x) dx T_{col}[t + a] + \epsilon_i$	(Equation 12).\\
These errors were independently sampled from the normal distribution with mean of 0.0 and SD of $\sigma$, and individually assigned to simulated $\rm{\delta^{15}N}$ values. By considering this individual error term, parameters that result in $D$ values smaller than $D_{opt}$ can be generated, which represent more plausible estimates for the measured data. In the ABC framework, $D$ values are calculated using randomly generated parameters from the prior distributions, then the parameters that result in $D$ values smaller than $D_{opt}$ become the posterior distributions.

\par The sequential Monte Carlo algorithm in our model proceeds as follows (see Sisson et al. \cite{Sisson2007} for more details):
\begin{enumerate}
	\item Set prior distributions $\pi(\cdot)$ for the parameters and the number of particles $j$ in one population. Calculate the final tolerance $\alpha_K$ (= $D_{opt}$) under the MLE framework and set decreasing tolerances. Set the population indicator $k$ = 1 (initialization).
	\item Set the particle indicator $j$ = 1 (initialization).
	\begin{enumerate}
		\item If $k$ = 1, independently sample $\theta^{**}$ from the prior distribution $\pi(\theta)$. If $k$ > 1, sample $\theta^*$ from the previous population ${\theta^{(i)}_{k-1}}$ with weights $W^{(i)}_{k-1}$, and perturb the particle to $\theta^{**}$ with transition kernel $\phi$. Simulate the change in the $\rm{\delta^{15}N}$ value $\delta^{15}N_{new}^{**}(t)$ with $\theta^{**}$ using equation 12. If $D^{**} - D_{opt} \geq \alpha_k$, $\theta^{**}$ are rejected and then repeat procedure 2(a).
	\item Set the indicators as follows:
	\begin{itemize}
		\item $\theta^{(j)}_k = \theta^{**}$,
		\item $W^{(j)}_k = 1$ (if $k = 1$), and
		\item $W^{(j)}_k = \frac{\pi(\theta^{(j)}_{k-1})}{\sum_{x = 1}^{J} W_{k-1}(\theta^{(x)}_{k-1}) \phi(\theta^{(j)}_k \mid \theta^{(x)}_{k-1})}$ (if $k > 1$).
	\end{itemize}
	If $j < J$, increment $j = j + 1$ and go to procedure 2(a).
	\end{enumerate}
	\item {Normalize the weights so that:\\
	$\sum_{j=1}^{J} W^{(j)}_k = 1$.\\
If the requirements for an effective sample size \textit{ESS} are not met such as:\\
	$ESS = \frac{1}{\sum_{j = J}^{J} {W^{(j)}_k}^2} < \frac{I}{2}$,\\
	sample with replacement, the particles $\theta^{(j)}_k$ with weights $W^{(j)}_k$ to obtain a new population $\theta^{(j)}_k$, and set weights $W^{(j)}_k = \frac{1}{J}$.}
	\item If $k < K$, increment $k = k + 1$ and go to procedure 2.
\end{enumerate}
Default prior distributions $\pi(\cdot)$ were set as normal distributions with default means of \{0.5, 3.0, 1.9, $\delta^{15}N_{mother}$, and 0.0\} and SDs of \{3.0, 3.0, 0.9, 3.0, and 1.0\} for {$t_{1}$, $t_{2}$, $E$, $\delta^{15}N_{wnfood}$, and $\sigma$}, respectively. The mean weaning age was obtained from values recommended by modern pediatricians and the biologically expected ages \cite{Dettwyler2004}. The mean and standard deviation of the enrichment factor $E$ was obtained from the values reported by Waters-Rist and Katzenberg \cite{Waters-Rist2010}. The hyper parameter for the individual error term $\sigma$ was used as an absolute value in the calculation. The default number of particles $J$ was 10000. Decreasing tolerances $\alpha_k$ were set as $D_{opt}$ + \{2, 1, 0.5, 0.25, 0.125, 0.0625, 0\} and, therfore, the number of populations $K$ = 7. The transition kernel $\phi$ was set to be a normal distribution with a mean of 0.0 and SD of 0.1.

\section*{Results and Discussion}
\subsection*{Subadult bone collagen turnover rate}
\par
\begin{figure}[!ht]
\begin{center}
	\includegraphics{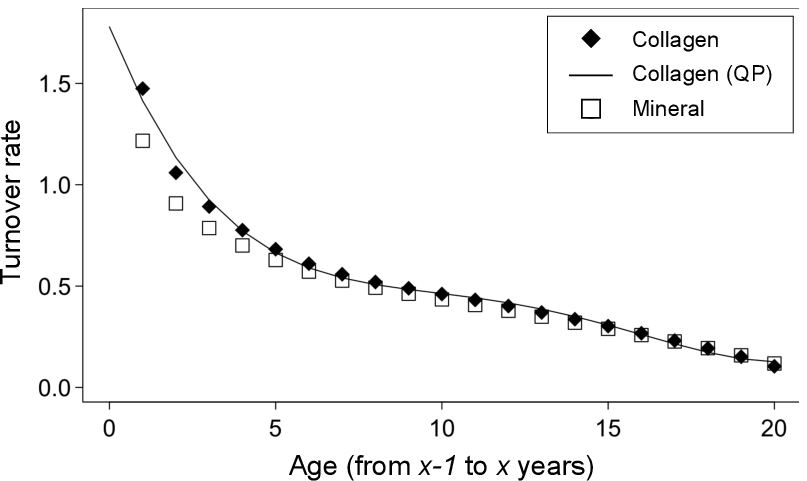}
\end{center}
\caption{{\bf Estimated temporal changes in bone mineral and collagen turnover rates.} Turnover rates of bone minerals and collagen are represented as discrete values, and that of collagen is fitted to QP plotted against age.}
\label{qp}
\end{figure}

 The calculated turnover rates are shown in Table \ref{turnover} and Figure \ref{qp}. The turnover rate of bone collagen was estimated to be larger than that of bone mineral until an individual reaches their late teens, and to decrease over the course of subadult growth. 

\begin{table}[!ht]
\caption{{\bf Estimated temporal changes in turnover rates for bone minerals and collagen.}}
\begin{center}
\begin{tabular}{l l l l l}
\hline
\multicolumn{2}{l}{Age} & \multicolumn{3}{l}{Turnover rate} \\
From & To & Mineral & Collagen & Collagen (QP) \\
\hline
0 & 1 & 1.217 & 1.474 & 1.413 \\
1 & 2 & 0.908 & 1.059 & 1.134 \\
2 & 3 & 0.786 & 0.892 & 0.924 \\
3 & 4 & 0.700 & 0.776 & 0.771 \\
4 & 5 & 0.629 & 0.682 & 0.664 \\
5 & 6 & 0.571 & 0.611 & 0.590 \\
6 & 7 & 0.527 & 0.558 & 0.540 \\
7 & 8 & 0.492 & 0.520 & 0.507 \\
8 & 9 & 0.462 & 0.489 & 0.483 \\
9 & 10 & 0.434 & 0.461 & 0.463 \\
10 & 11 & 0.407 & 0.432 & 0.441 \\
11 & 12 & 0.378 & 0.402 & 0.416 \\
12 & 13 & 0.349 & 0.370 & 0.386 \\
13 & 14 & 0.319 & 0.337 & 0.349 \\
14 & 15 & 0.289 & 0.302 & 0.306 \\
15 & 16 & 0.258 & 0.267 & 0.260 \\
16 & 17 & 0.227 & 0.231 & 0.213 \\
17 & 18 & 0.194 & 0.193 & 0.171 \\
18 & 19 & 0.158 & 0.151 & 0.139 \\
19 & 20 & 0.118 & 0.104 & 0.124 \\
\hline
\end{tabular}
\end{center}
\begin{flushleft}
	QP: calculated from the QP function.
\end{flushleft}
\label{turnover}
\end{table}

\par The integrated bone collagen turnover rate from 0.0 to 1.0 years of age was estimated to be 1.588, and the estimated bone collagen turnover rate was higher than 1.000 per year by two years of age (see Table \ref{turnover}). The integrated turnover rate from 0.0 years of age reached 0.966 at 0.60 years of age, suggesting that it takes 31 weeks for infants to fully reflect post-birth dietary $\rm{\delta^{15}N}$ signals. Tracer intake and biochemical marker studies have shown that the bone mineral and collagen turnover rates are high in the first few years of life (i.e., $>$1.0 per year) \cite{Bryant1964, Rivera1965, Szulc2000}, which is consistent with our results (see Table \ref{turnover} and Figure \ref{qp}). However, temporal changes in the bone collagen turnover rate after infancy and before adulthood have never been estimated directly and continuously, and the present study allowed them to be estimated. An isotopic study on an archaeological infant of a known age has suggested that infant rib bone collagen can fully reflect post-birth dietary $\rm{^{15}N}$ input, in an extreme case, in only five to six weeks \cite{Nitsch2011}, but this is estimated to take 31 weeks from our results. Our study allows typical temporal changes to be estimated, but the bone collagen turnover rate in subadulthood probably varies.

\par The integrated bone collagen turnover rate from 19.0 to 20.0 years of age was estimated to be 0.130 per year in our study (see Figure \ref{qp}), which is a little higher than that proposed by Stenhouse and Baxter (10.4 $\pm$ 2.7\% during adulthood, \cite{Stenhouse1979}) and Hedges et al. (9.7\% and 4.1\% for 20-year-old male and female femora, respectively, \cite{Hedges2007}). Although the type of bone sampled by Stenhouse and Baxter \cite{Stenhouse1979} is not stated, differences between the turnover rates in different bone types could cause these different results. The turnover rates are higher in bones with greater surface to volume ratios than those in bones with smaller ratios \cite{Parfitt2002}. Ribs, which were target bones in our study, have relatively high proportions of cancellous and thin cortical parts, whereas femur analyzed by Hedges et al. \cite{Hedges2007} has a lower proportion of cancellous and thick cortical parts. although there are slight differences, the overall trend of the temporal changes in bone turnover rates in this study is consistent with previous estimates.

\subsection*{The implemented model}
\par
\begin{figure}[!ht]
\begin{center}
	\includegraphics{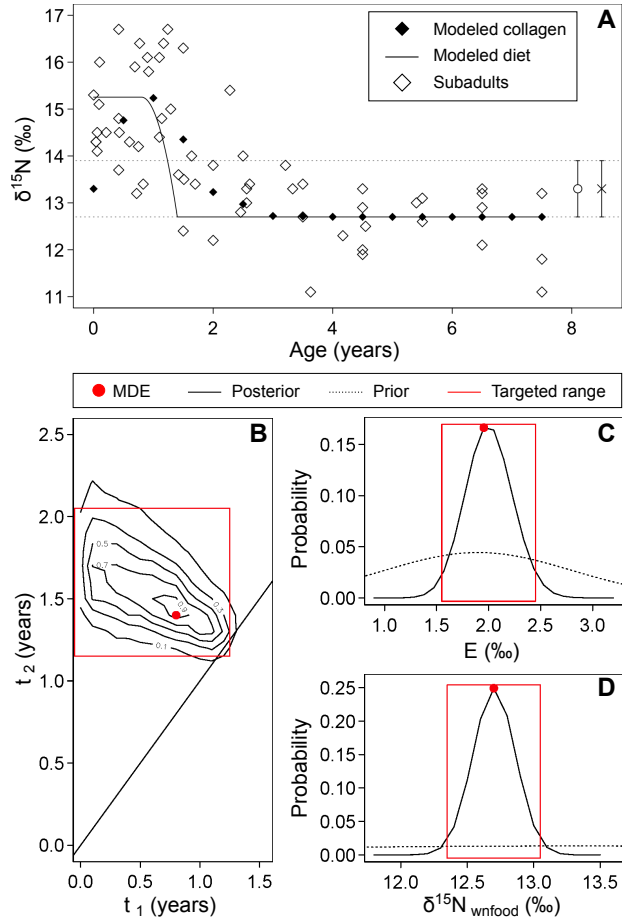}
\end{center}
\caption{{\bf An example of the results of applying WARN model using the Spitalfields population as a case study.} (A) Modeled temporal changes in the $\rm{\delta^{15}N}$ values by subadult age calculated from the reconstructed MDEs. Mean and SD ranges for adult females and all adults are indicated with open circles and crosses, respectively. (B) Contour lines show the posterior probability for the combination of weaning ages. The target ranges for $t_{1}$ and $t_{2}$ are 0.0--1.2 years and 1.2--2.0 years of age, respectively, and the calculated joint probability for the ranges is 0.956. (C) Distribution of posterior probabilities for the $\rm{^{15}N}$-enrichment from maternal to infant tissues. The target range is 1.6--2.4\textperthousand, and the calculated marginal probability for the range is 0.967. (D) Distribution of posterior probabilities for the $\rm{\delta^{15}N}$ values for collagen synthesized entirely from weaning foods. The target range is 12.4--13.0\textperthousand, and the calculated marginal probability for the range is 0.975. Subadult ages and bone collagen $\rm{\delta^{15}N}$ values were obtained from Nitsch et al. \cite{Nitsch2010, Nitsch2011}.}
\label{example}
\end{figure}

 The model developed in the present study is implemented as the R package WARN (Weaning Age Reconstruction with Nitrogen isotope analysis). Credible intervals can be calculated for a given parameter range using the WARN package. Images of the results calculated using the package are shown in Figure \ref{example}. Application of this model to previously reported skeletal populations and meta-analysis of the results will be reported elsewhere in the near future.

\par Although it is desirable to test the model validity, the absence of proper test data means this is not possible. Archaeological skeletal populations cannot be tested because the true weaning ages are usually unknowable, and historical literature, if any, describing breastfeeding practices at the time period when the population lived sometimes differs from actual practices (e.g. \cite{Fildes1982, Nitsch2011}; see also \cite{Dupras2001, Fuller2006a, Prowse2008}). Since the model presented here was intended for human subadult bones, conducting an experimental study was difficult, and hair, nail, and other tissues were not suitable for analysis because they have different turnover rates than bone collagen. Experimental studies of animals would not be appropriate because human growth patterns are unique among mammals \cite{Bogin1999, Kennedy2005}; therefore the nitrogen mass balance in human subadults would probably be different from that in other animals.

\par There are two caveats to consider before applying the model presented here. First, the present model is intended for bones with relatively high turnover rates, such as cancellous bones or ribs. Although WARN can be applied equally to isotopic data from bones with relatively low surface to volume ratios (e.g., limbs, cranium, and mandible), attention to this aspect is required for more precise analysis. Second, the WARN approach will always attempt to fit a model, even if the subadult $\rm{\delta^{15}N}$ values do not indicate breastfeeding and weaning signals. If researchers cannot find patterns of isotopic changes by visually inspecting the data, they are urged to examine their data carefully before applying the model, for example, for a biased age distribution or high isotopic variability in subadults. Although the estimated turnover rate and model developed can be further improved, in this study, we propose a framework for objectively and quantitatively analyzing and interpreting subadult bone collagen $\rm{\delta^{15}N}$ values. A precise reconstruction of past breastfeeding and weaning practices over a wide range of time periods and geographic regions could make it possible to understand this unique feature of human life history and cultural diversity in infant feeding practices \cite{Bogin1997, Hawkes1998, Humphrey2010, Kennedy2005, Lee1996, Sellen2007}.

\section*{Acknowledgments and Funding}
\par This study was supported in part by Grants-in-Aid for Scientific Research (KAKENHI: 24-785) from the Japan Society for the Promotion of Science.

\begin{minilinespace}

\end{minilinespace}

\end{document}